\documentclass[USenglish]{article}
\usepackage[utf8]{inputenc}
\usepackage[big,online]{dgruyter}

\usepackage{lmodern} 
\usepackage{microtype}
\usepackage[numbers,square,sort&compress]{natbib}
\setcitestyle{numbers,sort&compress}
\usepackage[cmyk]{xcolor}
\definecolor{blue}{cmyk}{1,0,0,0}

\theoremstyle{dgdef}

\usepackage{graphicx}
\usepackage{subfigure}
\usepackage{verbatim}
\usepackage{dcolumn}
\usepackage{bm}
\usepackage{epsf}
\usepackage{color}
\usepackage[toc]{appendix}
\usepackage{stmaryrd}
\usepackage{amsthm}
\usepackage{hhline}
\usepackage{amssymb}
\usepackage{tikz}
\usetikzlibrary{arrows,shapes,calc,matrix}

\expandafter\let\csname equation*\endcsname\relax
\expandafter\let\csname endequation*\endcsname\relax
\usepackage{amsmath}
\usepackage{xstring}

\makeatletter
\newcommand\footnoteref[1]{\protected@xdef\@thefnmark{\ref{#1}}\@footnotemark}
\makeatother

\newcommand{\e}[1]{\exp{\left(#1\right)}}

\newcommand{\bla}{bla\\bla\\bla\\bla\\bla}
\usepackage{fmtcount}

\newcommand{\mrm}[1]{\mathrm{#1}}

\renewcommand{\appendix}{
}

\newcommand{\draftmode}{1}    
\newcommand{\notetoself}[1]{\ifnum \draftmode=1 {\color[rgb]{0,0,0.8} [#1]} \fi}  
\newcommand{\cuttext}[1]{\ifnum \draftmode=1 {\color[rgb]{0,0.5,0} [#1]} \fi}  
\newcommand{\warntext}[1]{\ifnum \draftmode=1 {\color[rgb]{0.9,0.6,0} #1} \else {#1} \color{black} \fi}
\newcommand{\aref}[1]{{Appendix~\hyperref[#1]{A}}}
\newcommand{\bref}[1]{{Appendix~\hyperref[#1]{B}}}

\begin{document}

	\articletype{Research Article}

\title{Endoreversible Otto engines at maximal power}
\runningtitle{Endoreversible Otto engines at maximal power}

\author[1]{Zackary Smith}
\author[1]{P. S. Pal}
\author*[2]{Sebastian Deffner} 
\runningauthor{Z. Smith, P. S. Pal, and S. Deffner}
\affil[1]{\protect\raggedright 
Department of Physics, University of Maryland, Baltimore County, Baltimore, MD 21250, USA}
\affil[2]{\protect\raggedright 
Department of Physics, University of Maryland, Baltimore County, Baltimore, MD 21250, USA , email: deffner@umbc.edu}
	
	
\abstract{Despite its idealizations, thermodynamics has proven its power as a predictive theory for practical applications. In particular, the Curzon-Ahlborn efficiency provides a benchmark for any real engine operating at maximal power. Here we further develop the analysis of endoreversible Otto engines. For a generic class of working mediums, whose internal energy is proportional to some power of the temperature, we find that no engine can achieve the Carnot efficiency at finite power. However, we also find that for the specific example of photonic engines the efficiency at maximal power is larger than the Curzon-Ahlborn efficiency.}

\keywords{Quantum Thermodynamics, Optimal Heat Engines, Endoreversible Otto Cycle}

\maketitle

\section{Introduction}

Thermodynamics was invented in the beginning of the Industrial Revolution to understand and improve steam engines \cite{Kondepudi1998}. Rather curiously, however, in classic thermodynamics the only engine cycles that are fully described are comprised of idealized processes, that can be thought of as successions of equilibrium states \cite{Callen1985}. Thus, such ideal modes of operation of any engine have to be quasistatic, i.e., infinitely slow and the power output is strictly zero. In all practical applications, however, one is rather interested in the opposite, namely operating the engine at its maximal power output.

Building on the framework of endoreversible thermodynamics \cite{Hoffmann1997}, Curzon and Ahlborn spearheaded the study of engines cycles at maximal power. In endoreversible thermodynamics one assumes that all processes are slow enough that the system \emph{locally equilibrates}, yet the processes are too fast for the system to reach a state of equilibrium with the environment. More specifically, imagine an engine, whose working medium is at equilibrium at temperature $T$. However, $T$ is not equal to the temperature of the heat bath, $T_\mrm{bath}$, and thus there is a temperature gradient at the boundaries of the engine. Now further imagine that the engine undergoes a slow, cyclic state transformation, where slow means that the working medium remains \emph{locally} in equilibrium at all times. Then, from the point of view of the environment the device undergoes an irreversible cycle. Such state transformations are called \emph{endoreversible} \cite{Hoffmann1997}, which means that locally the transformation is reversible, but globally irreversible. 

Curzon and Ahlborn showed \cite{Curzon1975} that the efficiency of a Carnot engine undergoing an endoreversible cycle at maximal power is given by,
\begin{equation}
\label{eq:CA}
\eta_\mrm{CA}=1-\sqrt{\frac{T_c}{T_h}}\,,
\end{equation}
where $T_c$ and $T_h$ are the temperatures of the cold and hot reservoirs, respectively. Since its discovery the Curzon-Ahlborn efficiency \eqref{eq:CA} has received a great deal of attention. For instance, it has been found that also endoreversible Stirling cycles at maximum power are described by $\eta_\mrm{CA}$ \cite{Erbay1997}. Yet, it has also been shown that whether or not a finite time Carnot cycle really assumes $\eta_\mrm{CA}$ is determined by the ``symmetry'' of the dissipation \cite{Esposito2010}.

In the present analysis we focus on the Otto cycle. For maximally powerful Otto engines the situation is a little more involved, and it has been shown, e.g., that an endoreversible Otto engine with an ideal gas as working medium \cite{Leff1987} at maximal work does assume $\eta_\mrm{CA}$. Similarly, harmonic quantum Otto engines reach the Curzon-Ahlborn efficiency in the quasistatic limit  \cite{Rezek2006,Abah2012}. More recently, it was demonstrated that the efficiency of endoreversible Otto engines at maximal power depends on the equation of state, i.e., the nature of the working medium \cite{Uzdin2014,Deffner2018Entropy,Deffner2019book,Kloc2019,Myers2020} and the specific implementation of the power stroke  \cite{Chen2019,Chen2019PRE,Bonanca2018,Abah2019,Lee2020}. 

In the following analysis we will generalize and extend the framework developed by one of the authors in Ref.~\cite{Deffner2018Entropy}. In particular, it was shown in Ref.~\cite{Deffner2018Entropy} that endoreversible Otto engines with working mediums whose internal energy is linear in temperature, such as ideal gases or classical harmonic oscillators, do operate at the Curzon-Ahlborn efficiency \eqref{eq:CA}. However, it was also shown that more complicated working mediums, such as the quantum harmonic oscillator, can exhibit an efficiency that is significantly larger. This raises the natural question, whether there is an optimal working medium for which the efficiency of an endoreversible Otto engine achieves even the Carnot efficiency. For instance, using different techniques  Refs.~\cite{Bonanca2018} and \cite{Raz2016} showed that in principle it is possible to construct finite-time engines that do achieve the Carnot efficiency at maximal power output. In the present analysis we will show that this is \emph{not} the case for a generic class possible working mediums in endoreversible Otto engines.

\section{Preliminaries: endoreversible Otto cycle}

We start by briefly reviewing the endoreversible Otto cycle that was put forward in Ref.~\cite{Deffner2018Entropy}. The ideal Otto cycle is a four-stroke cycle consisting of isentropic compression, isochoric heating, isentropic expansion, and ischoric cooling \cite{Callen1985}. The endoreversible generalization is then comprised of the following four strokes:

\textit{Isentropic compression (reversible).} During isentropic processes thermodynamic systems do not exchange heat with the environment. Therefore, during the isentropic compression the working substance can be considered independent of the environment, and thus the stroke remains fully reversible. From the first law of thermodynamics, $\Delta E=Q+W$, we have,
\begin{equation}
\label{eq:Qcomp}
Q_\mrm{comp}=0\quad\mrm{and}\quad W_\mrm{comp}=E(T_2,V_2)-E(T_1,V_1)
\end{equation}
where $Q_\mrm{comp}$ is the heat exchanged, and $W_\mrm{comp}$ is the work performed during the isentropic compression from $V_1$ to $V_2$.

\textit{Isochoric heating (endoreversible).} During the isochoric strokes the volume, $V$, is held constant, and the system exchanges only heat with the environment. Thus, we have for isochoric heating 
\begin{equation}
\label{eq:Qheat}
Q_h=E(T_3,V_2)-E(T_2,V_2)\quad\mrm{and}\quad W_h=0\,.
\end{equation}
In complete analogy to Curzon and Ahlborn's original analysis \cite{Curzon1975} we now assume that the working substance is in a state of local equilibrium, but also that the working substance never fully equilibrates with the hot reservoir. Therefore, we can write
\begin{equation}
T(0)=T_2\quad\mrm{and}\quad T(\tau_h)=T_3\quad\mrm{with}\quad T_2<T_3 \leq T_h\,,
\end{equation}
where $\tau_h$ is the duration of the stroke.

Since the rate of heat flux cannot be assumed to be constant for isochoric processes \cite{Deffner2018Entropy} , we have to explicitly account for the change in temperature from $T_2$ to $T_3$. We have with the help of Fourier's law \cite{Callen1985} that,
\begin{equation}
\label{eq:fourier_hot}
\frac{d T}{dt}=-\alpha_h \left(T(t)-T_h\right)
\end{equation}
where $\alpha_h$ is a constant depending on the heat conductivity and heat capacity of the working substance. 

Equation~\eqref{eq:fourier_hot} can be solved exactly, and we obtain
\begin{equation}
\label{eq:rel_hot}
T_3-T_h=\left(T_2-T_h\right)\,\e{-\alpha_h \tau_h}\,.
\end{equation}
In the following, we will see that Eq.~\eqref{eq:rel_hot} is instrumental in reducing the number of free parameters.

\textit{Isentropic expansion (reversible).}  In complete analogy to the compression, we have for the isentropic  expansion,
\begin{equation}
\label{eq:Qexp}
Q_\mrm{exp}=0\quad\mrm{and}\quad W_\mrm{exp}=E(T_4,V_1)-E(T_3,V_2)\,.
\end{equation}

\textit{Isochoric cooling (endoreversible).} Heat and work during the isochoric cooling read,
\begin{equation}
\label{eq:Qcool}
Q_c=E(T_1,V_1)-E(T_4,V_1)\quad\mrm{and}\quad W_c=0\,,
\end{equation}
where we now have
\begin{equation}
T(0)=T_4\quad\mrm{and}\quad T(\tau_c)=T_1\quad\mrm{with}\quad T_4>T_1 \geq T_c\,.
\end{equation}
Similarly to above \eqref{eq:fourier_hot} the heat transfer is described by Fourier's law
\begin{equation}
\label{eq:fourier_cold} 
\frac{d T}{dt}=-\alpha_c \left(T(t)-T_c\right)\,,
\end{equation}
where $\alpha_c$ is a constant characteristic for the cold stroke. From the solution of Eq.~\eqref{eq:fourier_cold} we now obtain
\begin{equation}
\label{eq:rel_cold}
T_1-T_c=\left(T_4-T_c\right)\,\e{-\alpha_c \tau_c}\,,
\end{equation}
which properly describes the decrease in temperature from $T_4$ back to $T_1$.

\textit{Endoreversible Otto efficiency.} The efficiency of any heat engine is defined as work performed by the engine divided by the heat absorbed from the hot reservoir \cite{Callen1985},
\begin{equation}
\label{eq:eta}
\eta=-\frac{W_\mrm{tot}}{Q_h}=-\frac{E(T_2,V_2)-E(T_1,V_1)+E(T_4,V_1)-E(T_3,V_2)}{E(T_3,V_2)-E(T_2,V_2)}\,,
\end{equation}
where we used Eqs.~\eqref{eq:Qcomp}, \eqref{eq:Qheat}, \eqref{eq:Qexp}, and \eqref{eq:Qcool}. To further compute $\eta$ we now need to specify the working medium.

\section{Otto efficiency for generic working mediums}

As mentioned above, it was shown in Ref.~\cite{Deffner2018Entropy} that for working substances whose internal energies are linear in temperature, $E\propto T$, the endoreversible Otto efficiency \eqref{eq:eta} becomes identical to $\eta_{CA}$ \eqref{eq:CA} at maximal power. This scenario includes the harmonic oscillator, the ideal gas, and also the van der Waals gas. However, it was also shown in Ref.~\cite{Deffner2018Entropy} that for nonlinear equations of state $\eta$ \eqref{eq:eta} assumes other values. 

\subsection{Polynomial fundamental relations}

Therefore, we now consider the generic case with fundamental relation,
\begin{equation}
\label{eq:en_gen}
E=\epsilon\, VT^{N+1}\quad\text{and}\quad S=\sigma\, VT^{N}\
\end{equation}
where (as usual) $S$ is the entropy, and $\epsilon$ and $\sigma$ are constants that are specific for the working medium. 

The latter expressions allow to find a relationship between volume and temperature. To this end, note that we have by construction of the Otto cycle
\begin{equation}
S(T_{1},V_1)=S(T_{2},V_2)\quad\text{and}\quad S(T_{3},V_2)=S(T_{4},V_1)\,.
\end{equation}
Using the explicit form of the entropy \eqref{eq:en_gen} we immediately obtain
\begin{equation}
\kappa=\left(\frac{T_3}{T_4}\right)^N=\left(\frac{T_2}{T_1}\right)^N\,,
\end{equation}
where we introduced the compression ratio $\kappa\equiv V_1/V_2$. It is then a simple exercise to show
\begin{equation}
\label{eq:eta_otto}
\eta=1-\kappa^{-1/N}\,,
\end{equation}
which is the Otto efficiency for any working medium with Eq.~\eqref{eq:en_gen} for some value of $N$. 

\textit{Efficiency at maximal power.} In complete analogy to Curzon and Ahlborn \cite{Curzon1975} one can now define the power output per cycle,
\begin{equation}
\label{eq:power}
P=\frac{Q_h+Q_c}{\gamma(\tau_h+\tau_c)}
\end{equation}
where we absorbed the duration of the ``working strokes'' into the pre-factor $\gamma$. Substituting the expression for the internal energy \eqref{eq:en_gen} and Eqs.~\eqref{eq:fourier_cold} and \eqref{eq:fourier_hot}, the power can be written as $P=P(\kappa,\tau_h,\tau_c,N)$, where we assumed the temperatures of the heat baths, $T_h$ and $T_c$, are given and constant. The task is now to maximize the power $P$ has a function of the stroke times, $\tau_h$ and $\tau_c$, and the compression ratio, $\kappa$. Furthermore, $N$ is an additional free parameter. Maximizing $P$ as a function of $N$ would then allow to identify the ``optimal'' working medium with the highest maximal power. However, there is no guarantee that the determined value of $N$ corresponds to an existing physical system.

\subsection{No optimal working medium with Carnot efficiency}

Furthermore, the optimization problem can also be turned around, and we now ask for which $N$ the power is maximized as a function of $\kappa$ such that the efficiency \eqref{eq:eta_otto} becomes identical to the Carnot efficiency. 

To this end, we compute the analytical expression of the power \eqref{eq:power}, which results in a rather lengthy expression, see also Ref.~\cite{Deffner2018Entropy}. To avoid clutter in the formulas and for the sake of simplicity, we hence assume that the stroke times are identical, $\tau_c=\tau_h=\tau$, and also that the heat conduction coefficients are the same, $\alpha_c=\alpha_h=\alpha$. Using the explicit form of the internal energy \eqref{eq:en_gen} and the endoreversible heating laws \eqref{eq:fourier_cold} and \eqref{eq:fourier_hot} we obtain
\begin{equation}
\label{eq:power_general}
P(\eta,\tau)=\frac{\epsilon V_2\,\eta}{2 \gamma \tau\left(1+\e{\alpha\tau}\right)^{N+1}}\,\left[\left(\e{\alpha\tau}\,T_h+\frac{T_c}{1-\eta}\right)^{N+1}-\left(T_h+\frac{\e{\alpha\tau}\,T_c}{1-\eta}\right)^{N+1}\right]\,,
\end{equation}
where we also used the Otto efficiency \eqref{eq:eta_otto}.

We immediately notice two striking consequences: (i) The endoreversible power output is a reasonably simple function of the efficiency \eqref{eq:eta_otto}. However, this does not mean that the trade-off between power and efficiency is trivial. And more importantly, (ii) if $\eta$ is identical to the Carnot efficiency, $\eta=1-T_c/T_h$, the power exactly vanishes. Thus, no thermodynamic system with a fundamental relation of the form \eqref{eq:en_gen} can achieve the Carnot efficiency at any finite power.

\textit{Efficiency at maximal power.} The remaining task is to determine the efficiency at maximal power. To this end, we simply need to determine the maximum of $P(\eta,\tau)$ \eqref{eq:power_general} as a function of $\eta$. For general $N$ this leads to having to solve an $N$th order polynomial. Therefore, we will conclude the analysis with an analytically solvable example.

\subsection{Example: photonic Otto engine}

To ground the discussion in reality and for pedagogical reasons, we now chose $N=3$, for which Eq.~\eqref{eq:en_gen} describes the photonic gas \cite{Callen1985}. Specifically, we have
\begin{equation}
E=\epsilon\, VT^4\quad\text{and}\quad S=\sigma\, VT^3\,
\end{equation}
where 
\begin{equation}
\epsilon=\frac{\pi^{2}k_{B}^{4}}{15c^{3}\hbar^{3}}\quad\text{and}\quad \sigma=\frac{4}{3}\,\epsilon\,.
\end{equation}
Accordingly, the endoreversible power \eqref{eq:power_general} becomes,
\begin{equation}
\label{eq:power_photonic}
P(\eta,\tau)=\frac{\epsilon V_2\,\eta}{2 \gamma \tau\left(1+\e{\alpha\tau}\right)^{4}}\,\left[\left(\e{\alpha\tau}\,T_h+\frac{T_c}{1-\eta}\right)^{4}-\left(T_h+\frac{\e{\alpha\tau}\,T_c}{1-\eta}\right)^{4}\right]\,,
\end{equation}
which we now maximize as a function of $\eta$ for a fixed value of $\alpha\tau$. Note that $\epsilon$, $V_2$, and $\gamma$ enter the expression for $P(\eta)$ in Eq.~\eqref{eq:power_photonic} only as pre-factors. Thus, the value of $\eta$, for which $P(\eta)$ takes a maximum, is independent of  $\epsilon$, $V_2$, and $\gamma$.

The first derivative of $P(\eta)$ is then a third order polynomial with only one real root, which determines the efficiency at maximal power. 
\begin{figure}
\centering
\includegraphics[width=.7\textwidth]{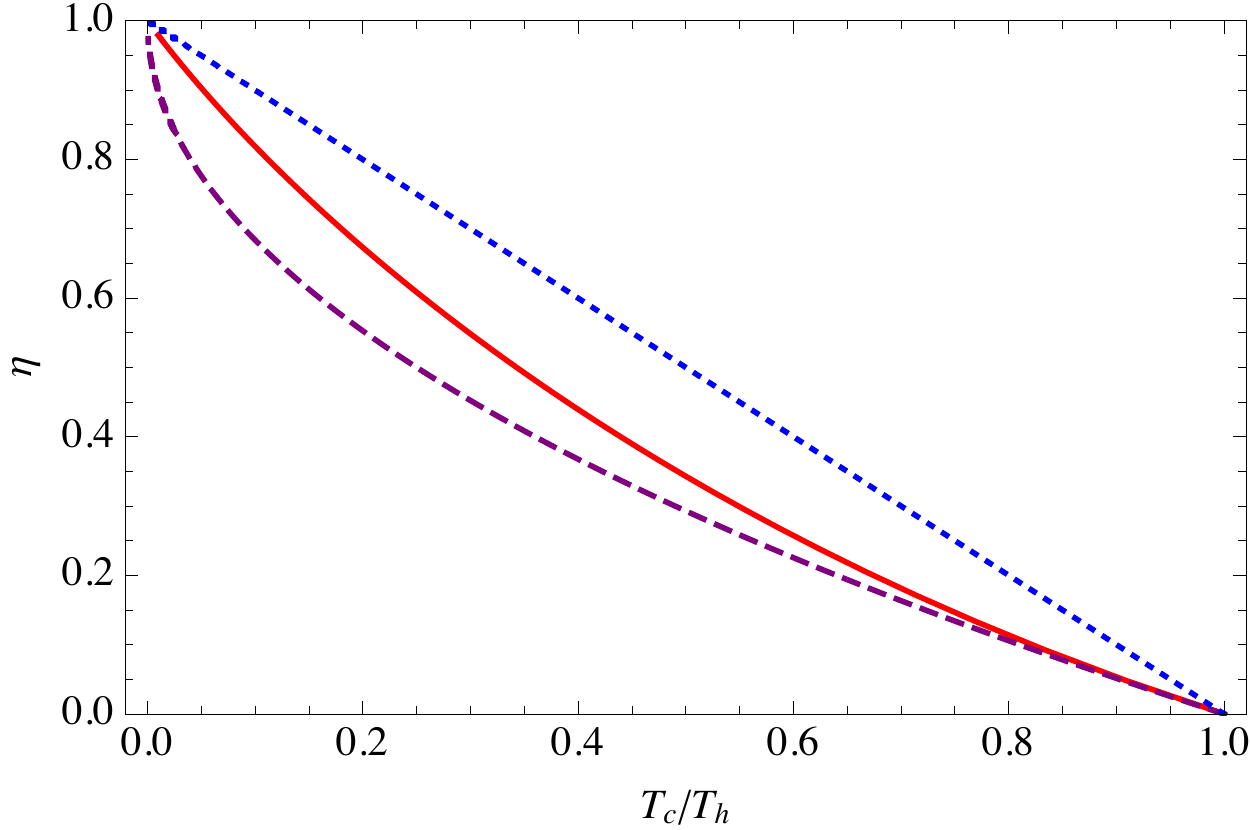}
\caption{\label{fig:photonic} Efficiency at maximal power (solid, red line) for the photonic Otto engine  \eqref{eq:power_photonic} together with the Curzon-Ahlborn efficiency (purple, dashed line) and the Carnot efficiency (blue, dotted line). Parameters are $\alpha\tau=1$.} 
\end{figure}
In Fig.~\ref{fig:photonic} we depict the solution for $\alpha\tau=1$. We observe that the efficiency is significantly larger than the Curzon-Ahlborn efficiency \eqref{eq:CA}, but also nowhere close the maximal efficiency at zero power, i.e., the Carnot efficiency.

\section{Concluding remarks}

In a previous work \cite{Deffner2018Entropy} it was shown that endoreversible Otto engines at maximal power do not necessarily operate at the Curzon-Ahlborn efficiency. In Ref.~\cite{Deffner2018Entropy} the focus was put on whether or not quantum effects could lead to an enhancement of the efficiency. However, it was found that it is actually the fundamental relation that determines the efficiency at maximal power. This observation was further solidified in the present analysis, in which we showed that systems whose internal energy is proportional to some power of the temperature can be treated analytically. We found that for this class of systems there is no ``optimal working medium'' for which finite power can be achieved at the Carnot efficiency. Finally, we showed that also for photonic gases the efficiency is larger than the Curzon-Ahlborn efficiency. This result is remarkable as this means that the efficiency of engines working with massless ideal gases (photons) is larger than the one working with massive ideal gases.

\begin{acknowledgement}
This work was conducted as part of the Undergraduate Research Program (Z.S.) in the Department of Physics at UMBC.
\end{acknowledgement}

\begin{funding}
This research was supported by grant number FQXi-RFP-1808 from the Foundational Questions Institute and Fetzer Franklin Fund, a donor advised fund of Silicon Valley Community Foundation.
\end{funding}

\bibliographystyle{apsrev4-1}
\bibliography{photonic_engine}

\end{document}